\documentclass[10pt]{iopart}
\usepackage{algorithmic}
\usepackage{graphicx}
\usepackage{textcomp}
\usepackage{multirow}
\usepackage{tabularx}
\usepackage{float}
\usepackage{stfloats}
\usepackage{subfigure}
\usepackage{cite}
\usepackage{hyperref}
\usepackage[hyphenbreaks]{breakurl}
\usepackage{array}
\begin{document}

\title{Cryogenic Characerization and Modeling of Standard CMOS down to Liquid Helium Temperature for Quantum Computing}

\author{Zhen Li$^{1,2}$,\ Chao Luo$^{1,2}$,\ Tengteng Lu$^{1,2}$,\ Jun Xu$^{2}$,\ Weicheng Kong$^{3}$,\ Guoping Guo$^{1}$ }
\address{$^1$Key Laboratory of Quantum Information, University of Science and Technology of China, Hefei, Anhui, 230026, China}
\address{$^2$Department of Physics, University of Science and Technology of China, Hefei, Anhui, 230026, China.}
\address{$^3$Origin Quantum Computing Company Limited, Hefei, Anhui, 230088, China}
\ead{gpguo@ustc.edu.cn}
\vspace{10pt}

\begin{abstract}
Cryogenic characterization and modeling of 0.18$\mu$m CMOS technology (1.8V and 5V) are presented in this paper. Several PMOS and NMOS transistors with different width to length ratios(W/L) were extensively characterized under various bias conditions at temperatures ranging from 300K down to 4.2K. We extracted their fundamental physical parameters and developed a compact model based on BSIM3V3. In addition to their I-V characteristics, threshold voltage(V$_{th}$) values, on/off current ratio, transconductance of the MOS transistors, and resistors on chips are measured at temperatures from 300K down to 4.2K.  A simple subcircuit was built to correct the kink effect. This work provides experimental
 evidence for implementation of cryogenic CMOS technology, a valid industrial tape-out process model, and promotes the application of integrated circuits in cryogenic environments, including quantum measurement and control systems for quantum chips at very low temperatures.
\end{abstract}

%
\vspace{2pc}
\noindent{\it Keywords}: Cryogenic electronics, MOSFETs, characterization, modeling, threshold voltage, kink effect, liquid helium temperature.
%
%
%
\ioptwocol

\section{Introduction}

\begin{figure}[htbp]
\centering
\includegraphics[width=\columnwidth]{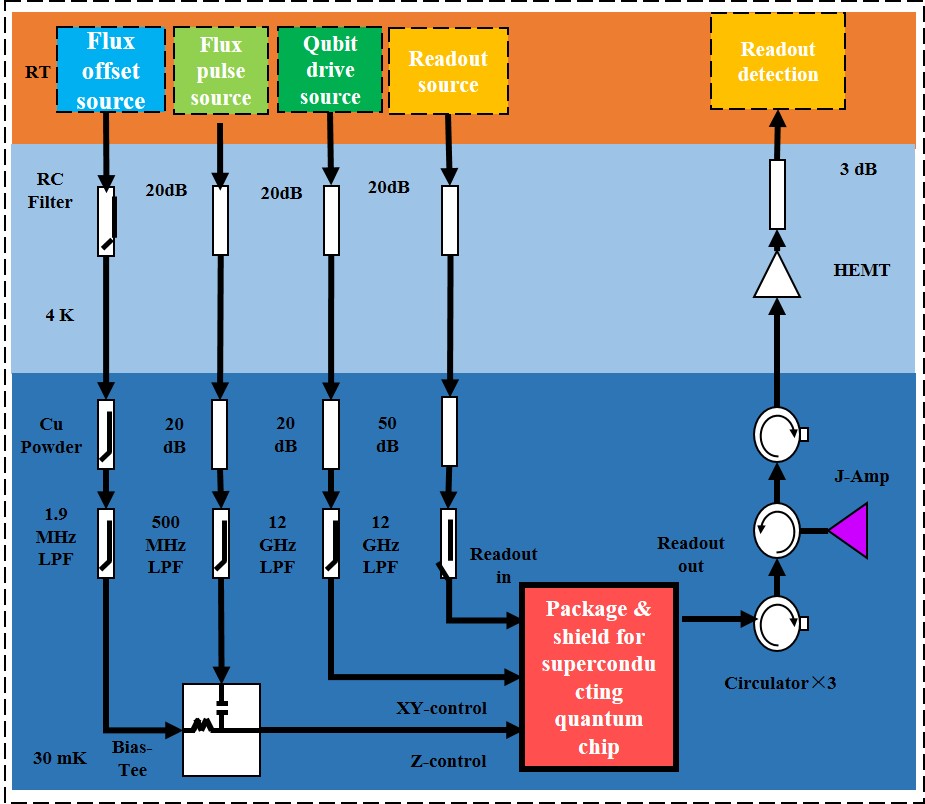}
\caption{Superconducting quantum chip readout-control system.}
\label{fig1}
\end{figure}

\begin{figure}[htbp]
\centering
\includegraphics[width=\columnwidth]{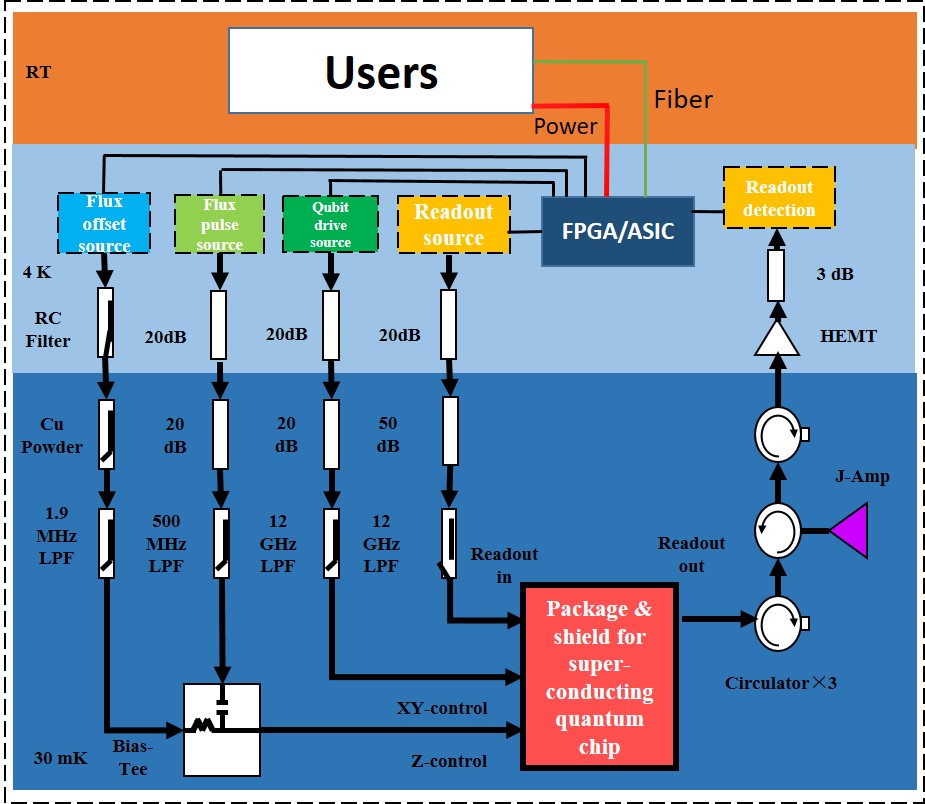}
\caption{Superconducting quantum chip readout-control system using Cryo-CMOS technology.}
\label{fig2}
\end{figure}

Cryogenic electronics have good prospects in application ranging from space exploration to infrared focal plane array\cite{7999541,604075,4089134}, and has been studied for use in quantum computing in recent years\cite{5373940,PhysRevApplied.3.024010,mine:isscc_2017_cryoCMOS,doi:10.1063/1.4979611}. A quantum computer comprises a quantum processor and a classical electronic control system\cite{8358026}. The quantum processor works in a dilution refrigerator at deep-cryogenic temperatures down to the milikelvin range(Fig. \ref{fig1}), while the electronic readout and control system is implemented using room-temperature (RT) laboratory instruments\cite{doi:10.1021/acs.nanolett.5b02400,PhysRevApplied.9.034011}. The requirements for wiring between the cryogenic quantum processor and the RT readout controller are becoming more expensive and less reliable as quantum chips become increasing complexed and highly integrated\cite{PhysRevApplied.3.024010}. Cryogenic complementary metal-oxide-semiconductor (Cryo-CMOS) technology can greatly reduce the thermal noise caused by non-ideal long signal lines and improve the signal-to-noise ratio and sensitivity of the quantum chip signals. Obtaining purer quantum control and readout signals with low delay efficiently improves quantum chip performance. Quantum Processors work in dilution refrigerators with maximum effective cooling powers of several hundreds $\mu$W. This constraint is relaxed at the liquid helium temperature (LHT) 4.2K, where moderate power dissipation is tolerable. Fig. \ref{fig2} shows the future quantum interface using Cryo-CMOS technology, we need to implement ADC (Analog Digital Converter), DAC (Digital Analog Converter), oscillator, FPGA (Field Programmable Gate Array), and other integrated circuits at cryogenic temperatures. Unfortunately Cryo-CMOS faces several challenges, including the power limitation of refrigerators, and interconnection, packaging and device modeling\cite{1484563,30958,1580601}.

The first problem to solve when designing Cryo-CMOS circuits is transistor modeling. SPICE model act as a bridge between device characterics and IC design. BSIM3\cite{Y.Cheng1995}, which is an industry-standard model, is valid from 230K to 430K for submicron processes. However, MOSFET characteristics changes at lower temperatures because of freeze-out effect, which has led to a requirement for SPICE model development for cryogenic temperatures\cite{47796,BALESTRA19941967,24362,BALESTRA1987321,3372,1052555}. Previous work has demonstrated that CMOS technologies have been characterized at temperatures down to 4K\cite{8269294,8066592,d0d52351b8e84fcc8e961be5fa144ba2}. However, BSIM model parameters have only been extracted down to 77K, and no systematic modeling of PMOS and NMOS devices with different width-to-length ratios are performed under different bias conditions at lower cryogenic temperatures, to the best of our knowledge\cite{7999541,ZHAO201449,1742-6596-834-1-012002,8329135,8424046}.

In this paper, characterization of SMIC 0.18 $\mu$m CMOS transistors and a compact SPICE model based on BSIM3v3\cite{Y.Cheng1995} are presented from 300K down to 4.2K. It is an aluminum interconnect process, compared with copper, aluminum has better electrical properties at low temperatures\cite{doi:10.1080/14786436708227694,doi:10.1063/1.555614}. Temperature-dependent parameters are revised at 4.2K and the model shows good agreement with measurement results. The 0.18 $\mu$m process V$_{th}$ and resistance of active area are measured from 300K to 4.2K for the first time.  This work is the first BSIM SPICE model range down to 4.2K for standard CMOS technology; the model can be applied directly to device and circuit electronic design automation(EDA) simulations.

\section{Measurement Setup}
\begin{figure}[htpb]
\centering
\includegraphics[width=0.7\columnwidth,height=0.7\columnwidth]{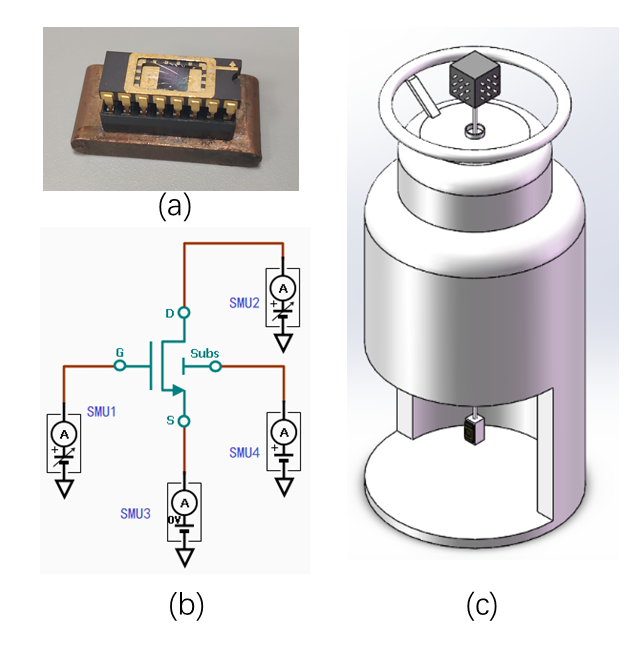}
\caption{(a): Al-bonded sample chip on a chip carrier. (b): Schematic of four-teminal MOS transistor DC measurment. (c): Schematic depiction of our cryogenic test setup. The sample is mounted on a long steel pipe which is shifted into a helium Dewar to reach the liquid helium. The connections to the sample are made with long (standard) cables that are attached to a break-out box on the top. There are 32 shielded connections to two 8$\times$2 DIP lock sockets, 4 DC connections to a resistor therometer.}
\label{fig3}
\end{figure}

Measurements of CMOS transistors with two different oxide thicknesses and a wide range of device sizes were performed, as shown in Table \ref{table1}. The sample chips were first pasted and wire bonded to chip-carriers using Al-wire bonds (Fig. \ref{fig3}(a)). These chip-carriers were then immersed in liquid nitrogen (77 K) and liquid helium (4.2 K) using a dipstick. A schematic of the cross-section of the setup is shown in Fig. \ref{fig3}(c). The dipstick consists of a 1.8m steel pipe with a break-out box for cables placed at the top end and two dual in-line package(DIP) lock sockets at the lower end of the pipe. In total, 36 cables (enamel insulated wire) are used for the DC connections, including four cables for the temperature sensor. The temperature sensor is a Rh-Fe thermometer with a 1.2K-325K measurement range. The cable resistance is 0.3-0.4$\Omega$, it is negligible compared with the resistance of MOSFET which is several hundreds or thousands ohms. Because the pipe can move up and down through a vacuum flange, the temperature can be shifted from 4.2K in the liquid phase to approximately 250K in the helium vapour at the top of the Dewar.

\begin{table}[htbp]
  \centering
  \footnotesize
    \setlength{\tabcolsep}{5pt}
\caption{SUMMARY OF CHARACTERIZED DEVICES}
    \begin{tabular}{ccccc}
    \br
    Technology & \multicolumn{4}{c}{ SMIC 0.18um Bulk CMOS Process} \\
    \mr
    Oxide & \multicolumn{2}{c}{Thin(3.6nm)} & \multicolumn{2}{c}{Thick(11.9nm)} \\
    \mr
    Norminal Voltage & \multicolumn{2}{c}{1.8V} & \multicolumn{2}{c}{5V} \\
   \mr
    Type  & NMOS  & PMOS  & NMOS  & PMOS \\
    \mr
    \multirow{7}[0]{*}{W/L[um/um]} & 100/0.18 & 100/0.18 & 100/0.6 & 100/0.5 \\
          & 10/10 & 10/10 & 10/10 & 10/10 \\
          & 10/0.6 & 10/0.6 & 10/2  & 10/2 \\
          & 10/0.2 & 10/0.2 & 10/0.65 & 10/0.55 \\
          & 10/0.18 & 10/0.18 & 10/0.6 & 10/0.5 \\
          & 10/0.16 & 10/0.16 & 10/0.5 & 10/0.45 \\
          & 0.22/0.18 & 0.22/0.18 & 0.3/0.6 & 0.3/0.5 \\
          \br
    \end{tabular}%
  \label{table1}%
\end{table}%

All the MOSFET electrical measurements were performed using a Keysight B1500A semiconductor device analyzer, as shown in Fig. \ref{fig3}(b). For the thin-oxide NMOS, we measured transfer characteristics in both linear (drain-source voltage V$_{DS}$ = 50 mV) and the saturation regions (V$_{DS}$ = 1.8V) under various substrate bias voltages, along with the output characteristics under zero substrate bias(bulk-sourse voltageV$_{BS}$=0V) and reverse bias voltage(V$_{BS}$=-1.8V) for various gate voltages(V$_{GS}$). For the thick-oxide MOS, bias condition were increased to 5V, while for PMOS, the bias conditions were reversed. The resistance measurements were performed using a Keysight 3458A Digital Multimeter with 8$\frac{1}{2}$ Digit precision.

\section{Characterization}

\begin{figure*}[htbp]
\centering
\includegraphics[width=0.95\textwidth]{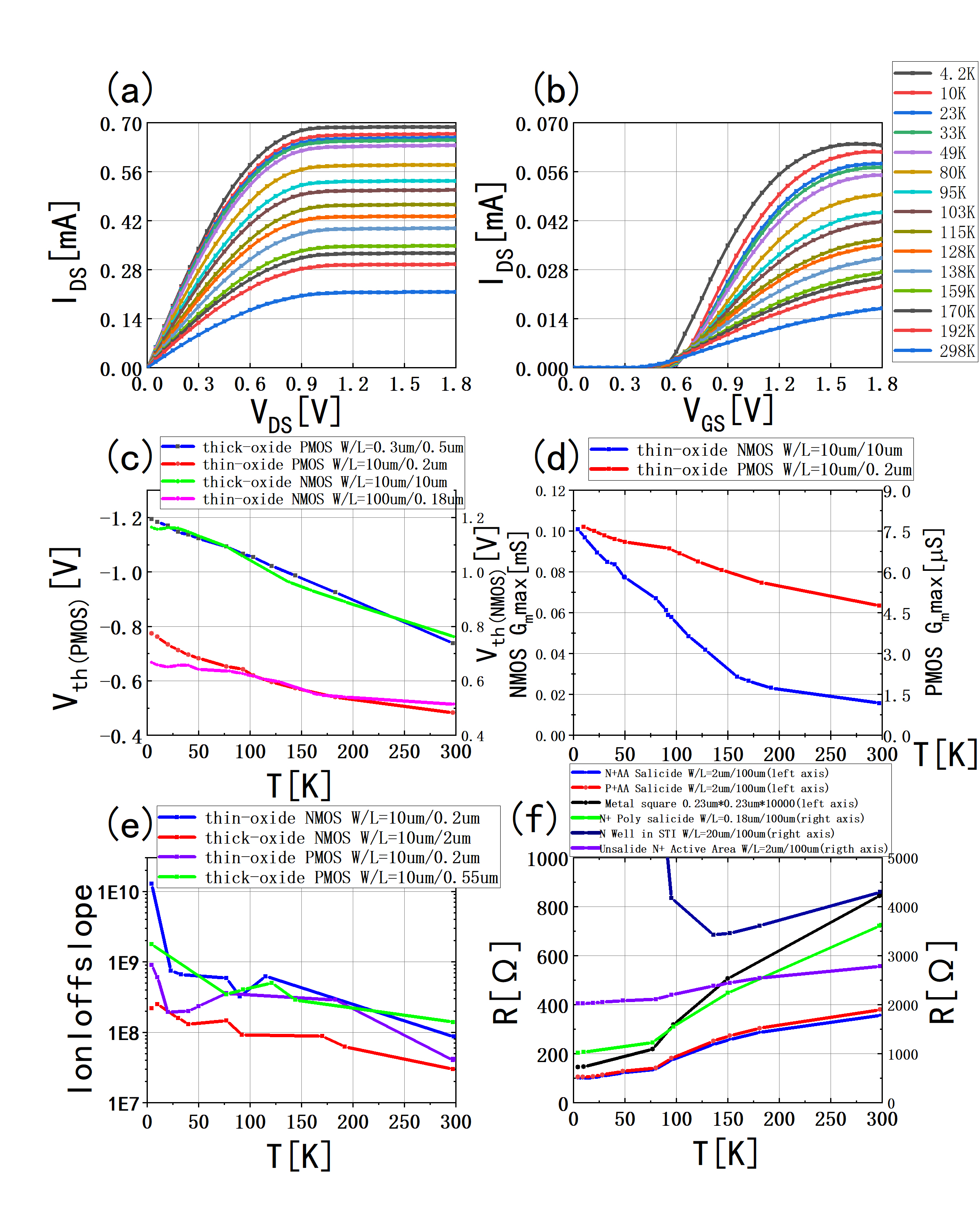}
\caption{(a): I$_{DS}$-V$_{DS}$ curves at different cryogenic temperatures of thin-oxide NMOS,V$_{DS}$=0V$\rightarrow$1.8V, W/L=10$\mu$m/10$\mu$m, V$_{GS}$=1.8V, V$_{BS}$=0V. (b): I$_{DS}$-V$_{GS}$ curves at different cryogenic temperatures of thin-oxide NMOS,V$_{GS}$=0V$\rightarrow$1.8V, W/L=10$\mu$m/10$\mu$m, V$_{DS}$=0.05V, V$_{BS}$=0V. (c): V$_{th}$ of CMOS transistors at different cryogenic temperatures. (d): Ion to Ioff ratio of CMOS transistors at different temperatures. (e): G$_{m}$max of MOSFETs at different cryogenic temperatures. (f): resistors on chip at different cryogenic temperatures, the six lines from top to bottom are N++ salicide area, W/L=2$\mu$m/100$\mu$m; P++ salicide area, W/L=2$\mu$m/100$\mu$m; Al metal square, 0.23$\mu$m$\ast$0.23$\mu$m$\ast$10000; N+ poly salicide, W/L=0.18$\mu$m/100$\mu$m; N Well in STI, W/L=20$\mu$m/100$\mu$m; Unsalicide N+, W/L=2$\mu$m/100$\mu$m }
\label{fig4}
\end{figure*}

The characteristics of CMOS transistors at different cryogenic temperatures are  shown in Fig. \ref{fig4}. As shown in  Fig. \ref{fig4}(b) and Fig. \ref{fig4}(c), threshold voltage increases as temperature decreases because of carrier freeze-out in the MOSFET channel region and thus a higher gate drive voltage is required to inject carriers into the channel region. V$_{th}$ varies approximately linearly with temperature, especially in the PMOS. Impurity freeze-out becomes important for temperature lower than 150K for shallow-energy-level dopants. At liquid nitrogen temperature, weak freeze-out takes place, which is mainly annoying for lightly doped drain (LDD) devices. At very low temperatures ($<$10 K), donor or acceptor impurities currently used to dope the semiconductor are fully frozen-out, at liquid helium temperature, practically no carriers remain in the bands if no field is applied\cite{24362,BALESTRA1987321,3372,1052555}.

Fig. \ref{fig4}(a) shows that I$_{DS}$ increases as temperature decreases, because the series resistance decreases as shown in Fig. \ref{fig4}(f), and mobility increase at cryogenic temperatures\cite{30958}. In low impurity concentration situations, such as p-well, the resistivity decreases due to an increase of mobility down to liquid nitrogen temperature\cite{ZHAO201449}. However, carrier freeze-out causes a steep increase in n-well resistivity at liquid helium temperature. Other resistances drop with decreasing temperature down to liquid helium temperature. The resistance of the salicide's heavy-doped N/P active area is reduced by 2/3 when compared with the RT value. The resistance of the unsalicide's heavy-doped N active area is reduced by 1/3 when compared with the RT value. In particular, Aluminum resistance drops drastically because of reduced lattice vibrations. This gives a great advantage to reducing noise caused by CMOS switching and power-supply line resistances. This is an advantage of Cryo-CMOS technology compared with RT CMOS technology.

\begin{figure*}[htpb]
\centering
\includegraphics[width=\textwidth]{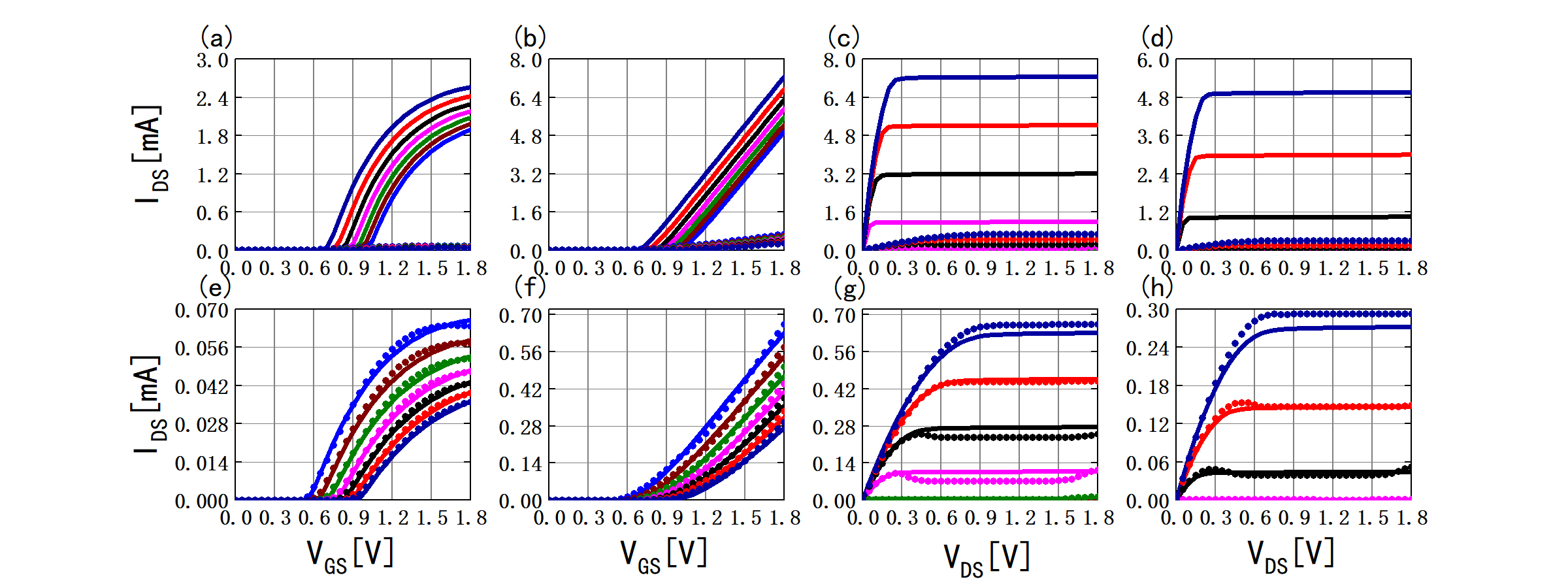}
\caption{I$_{DS}$-V$_{GS}$ curves (a,b,e,f) and I$_{DS}$-V$_{DS}$ curves (c,d,g,h) of thin-oxide NMOS at LHT. Device size (W/L) is 10$\mu$m/10$\mu$m.(a)-(d):before extraction, (e)-(h):extraction results, measured data: dashed lines; simulated data: solid lines.
(a),(e): V$_{BS}$=0V$\rightarrow$-1.8V, V$_{DS}$=0.05V; (b),(f): V$_{BS}$=0V$\rightarrow$-1.8V, V$_{DS}$=1.8V; (c),(g): V$_{GS}$=0V$\rightarrow$1.8V, V$_{BS}$=0V; (d),(h): V$_{GS}$=0V$\rightarrow$1.8V, V$_{BS}$=-1.8V.}
\label{fig5}
\end{figure*}

\begin{figure*}[htpb]
\centering
\includegraphics[width=\textwidth]{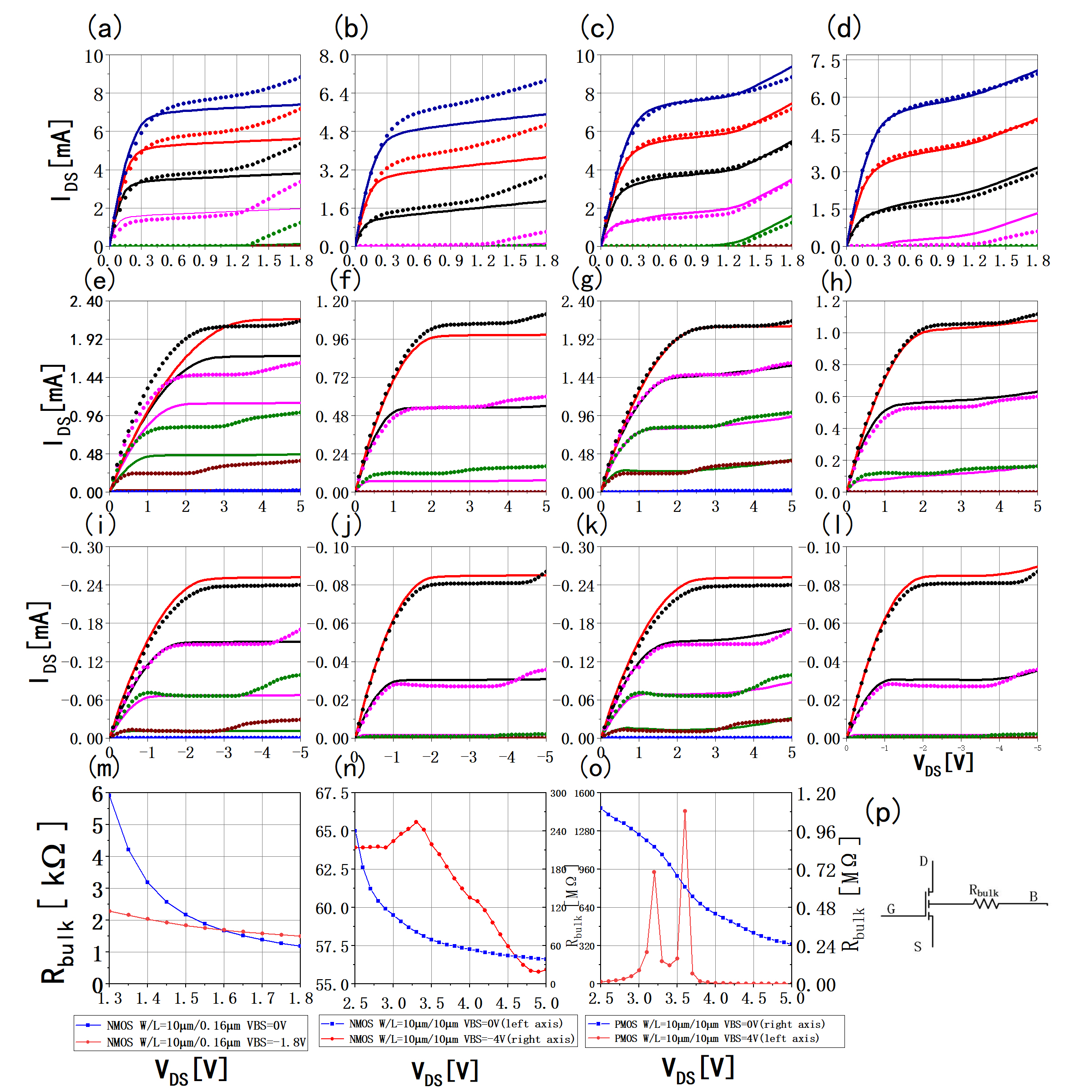}
\caption{Cryogenic kink correction.(a)-(l): I$_{DS}$-V$_{DS}$ curves of CMOS at LHT. Device size (W/L):(a)-(d) thin-oxide NMOS 10$\mu$m/0.16$\mu$m,(e)-(h) thick-oxide NMOS 10$\mu$m/10$\mu$m,(i)-(l) thick-oxide NMOS 10$\mu$m/10$\mu$m. (a)-(b),(e)-(f),(i)-(j):before kink correction, (c)-(d),(g)-(h),(k)-(l):kink correction results. measured data: dashed lines, simulated data: solid lines.
(a),(c): V$_{GS}$=0V$\rightarrow$1.8V, V$_{BS}$=0V; (b),(d): V$_{GS}$=0V$\rightarrow$1.8V, V$_{BS}$=-1.8V; (e),(g): V$_{GS}$=0V$\rightarrow$5V, V$_{BS}$=0V; (f),(h): V$_{GS}$=0V$\rightarrow$5V, V$_{BS}$=-4V; (i),(k): V$_{GS}$=0V$\rightarrow$-5V, V$_{BS}$=0V; (j),(l): V$_{GS}$=0V$\rightarrow$-5V, V$_{BS}$=4V;
(m) Sub-circuit resistance of thin-oxide MOSFET versus V$_{DS}$ at LHT, W/L=10$\mu$m /0.16$\mu$m.(n) Sub-circuit resistance of thick-oxide NMOSFET versus V$_{DS}$ at LHT, W/L=10$\mu$m /10$\mu$m.(o) Sub-circuit resistance of thick-oxide PMOSFET versus V$_{DS}$ at LHT, W/L=10$\mu$m /10$\mu$m.
(p)Schematic representation of sub-circuit model. }
\label{fig6}
\end{figure*}

The kink effect at the LHT is shown in Fig.\ref{fig6} and is caused by the LDD freezing out and substrate freeze-out. Impurity ionization will decrease as temperature decreases, especially in the light-doped regions near the source and drain. When the source-drain voltage becomes very high, impurity ionization will be activated to restrain the freeze-out effect under strong electric fields, CMOS transistors turn on for the second time\cite{BALESTRA19941967}. This freezing effect causes the source-drain parasitic resistance to decrease and then turn to normal. The second reason is the substrate freeze-out, since at very low temperatures the MOS structure has a type of floating substrate potential within the depletion region. Although the applied substrate voltage on the backside of the device is fixed, the depletion region is in a floating state. In this structure, the majority carrier current (substrate current) cannot reach the substrate contact and thus flows through the substrate to the source. Due to the increase of the majority carrier current with increasing drain voltage, flowing through the substrate to the source at increasing drain voltage, this substrate potential within the depletion region increases and causes a decrease of the threshold voltage, for sufficient drain voltage. Therefore, at very low temperature, we obtain an excess drain current which creates a kink in the current-drain voltage characteristics\cite{BALESTRA1987321}. Additionally, the measured saturation current value is less than the simulated value of BSIM3v3 model because of the channel freeze-out effect(Fig.\ref{fig5}(a-d)).

Ion to Ioff ratio(turn-on current and turn-off current ratio) is a typical parameter for MOSFET in digital integrated cirtuits. Ion to Ioff ratio maintains high value at cryogenic temperatures as shown in Fig. \ref{fig4}(d). The standard process CMOS can work well as a switch at low temperatures for digital circuits with low static power consumption, which is of great significance to the limited cooling power. Gate transconductance(G$_{m}$) indicates the gate-to-source current control capability. G$_{m}$max increases when temperature increases(Fig. \ref{fig4}(e)), increase of G$_{m}$ will supply wider bandwidth for the same power budget. Hence, MOS can work in analog circuits if it is modeling accurated.

\section{Modeling}

\begin{table*}[htbp]
  \centering
  \footnotesize
  \setlength{\tabcolsep}{1pt}
  \caption{MODEL PARAMETERS}
    \begin{tabular}{ccccccccccccc}
    \br
    Oxide & \multicolumn{6}{c}{Thin(3.6nm)} & \multicolumn{6}{c}{Thick(11.9nm)} \\
    \mr
    Norminal Voltage & \multicolumn{6}{c}{1.8V} & \multicolumn{6}{c}{5V} \\
    \mr
    Type & \multicolumn{3}{c}{NMOS} & \multicolumn{3}{c}{PMOS} & \multicolumn{3}{c}{NMOS} & \multicolumn{3}{c}{PMOS} \\
     \mr
    Temperature[K] & 300K & 77K & 4.2K & 300K & 77K & 4.2K & 300K & 77K & 4.2K & 300K & 77K & 4.2K \\
     \mr
    Vth0[V] & 0.39 & 0.4671 & 0.296 & -0.607 & -0.5785 & -0.4598 & 0.724 & 1.1391 & 0.31124 & -0.834 & -1.2 & -1.0532 \\
    K1[V$^{\frac{1}{2}}$] & 0.680104 & 0.53197 & 0.6966 & 0.87354 & 0.89101 & 0.8821 & 0.97 & 0.84906 & 1.6667 & 0.986 & 0.77845 & 1.1863 \\
    K2 & -0.04998 & -0.00013 & -0.00013 & -0.04666 & -0.05506 & -0.0534 & -0.015 & -0.00686 & -0.01878 & 0.0626 & 0.096653 & 0.035749 \\
    u0[cm$^2$/V$\cdot$s] & 340 & 606.51 & 116.09 & 0.0085 & 0.015369 & 0.000933 & 0.04415 & 0.36688 & 0.00426 & 0.015 & 0.076605 & 0.0035 \\
    ua[m/V] & -1E-09 & -4.4E-12 & -4.5E-12 & 2.5E-10 & 2.4E-10 & 3.52E-10 & -3.7E-10 & 1.11E-10 & 3.79E-10 & 2.27E-09 & 8.07E-09 & 4.84E-09 \\
    ub[(m/V)$^2$] & 0.236667 & 1.12E-17 & 6.46E-18 & 9.29E-19 & 9.29E-19 & 8.37E-19 & 2.6E-18 & 5.55E-18 & 5.65E-18 & 2.51E-20 & -3.6E-21 & -4.5E-19 \\
    uc[1/V] & 1.2E-10 & 3.33E-13 & 3.33E-13 & -7.2E-11 & -1E-10 & -1.7E-10 & 8.38E-11 & 8.48E-12 & 1E-12 & -7.6E-11 & -3E-10 & -4.9E-10 \\
    Dvt0 & 1.3 & 0.2165 & 5 & 1.03 & 3.0099 & 4.5143 & 7.46 & 7.46 & 7.46 & 2.8982 & 2.0023 & 26.374 \\
    Dvt1 & 0.577164 & 2.3925 & 5 & 0.35 & 0.48288 & 0.14103 & 0.805 & 0.805 & 0.805 & 0.6 & 0.21882 & 0.46447 \\
    Dvt2[1/V] & -0.17176 & -0.01816 & -3 & 0.052 & 0.000804 & 0.000861 & 0 & 0 & 0 & 0 & -0.047 & -0.6 \\
    Vsat[m/sec] & 82500 & 89809 & 53820 & 109320 & 4202600 & 26162 & 77500 & 77500 & 77500 & 88000 & 47264 & 98296 \\
    a0 & 0.83 & 0.70733 & 1.1883 & 1.0951 & 0.87389 & 0.83893 & 0.96 & 0.96 & 0.013849 & 1 & 0.73796 & 0.5917 \\
    ags[1/V] & 0.32 & 0.73655 & 0.1485 & 0.3024 & 0.39778 & 0.65873 & 0.15 & 0.15 & 0.036746 & 0.0655 & 0.006383 & 0.002047 \\
    Keta[1/V] & -0.003 & -3.6E-05 & -3.6E-05 & -0.0389 & -0.03197 & 0.05346 & -0.0015 & -0.0015 & -0.00169 & -0.005 & -0.00094 & -0.013 \\
    \br
    \end{tabular}%
  \label{table2}%
\end{table*}%

\begin{table*}[htbp]
  \centering
  \footnotesize
  \caption{RMS ERROR}
    \begin{tabular}{lcccccc}
    \br
    &\centre{3}{IDS(VDS)(VBS=0V) } & \centre{3}{IDS(VDS)(VBS=VBB$^{a}$) }\\
   \ns
      & \crule{6}\\
     &\multicolumn{1}{l}{Default} & \multicolumn{1}{l}{Revised} & \multicolumn{1}{l}{Corrected} & \multicolumn{1}{l}{Default} & \multicolumn{1}{l}{Revised} & \multicolumn{1}{l}{Corrected} \\
      \mr

    thin-oxide NMOS W/L=10$\mu$m/0.16$\mu$m & 13.37\% & 5.74\% & 1.71\% & 8.94\% & 5.71\% & 2.41\% \\
    thick-oxide NMOS W/L=10$\mu$m/10$\mu$m & 25.72\% & 5.77\% & 4.66\% & 27.54\% & 6.12\% & 1.67\% \\
    thick-oxide PMOS W/L=10$\mu$m/10$\mu$m & 174.15\% & 4.41\% & 1.41\% & 58.19\% & 3.25\% & 0.96\% \\
    \br
    \end{tabular}%

   $^{a}$for thin-oxide NMOS,VBB=-1.8V; for thick-oxide NMOS, VBB=-4V; for thick-oxide PMOS, VBB=4V.
  \label{table3}%
\end{table*}%

BSIM3 is a semi-empirical but accurate compact model for submicron process. We continue to adopt BSIM's mode and use semi-empirical methods to modeling CMOS device at very low temperatures. Our object is to extend the BSIM model in submicron process to cryogenic temperatures, focus on the accuracy of the model. Fitting errors between simulation with default parameters and LHT measured data are generally higher than 60\% under all bias conditions(Fig. \ref{fig5}(a-d)), while the BSIM3v3 model achieves a good degree of fitting at RT. First, the extraction procedure is performed using BSIMProPlus. The extraction process performed in the following procedure is based on physical understanding of the model and local optimization. The procedure can be described as follows:1) extract threshold voltage parameters such as V$_{th0}$, K1, K2 through large size device (large W$\&$L); 2) extract carrier mobility parameters such as $\mu$0, $\mu$a, $\mu$b, $\mu$c through large size device (large W$\&$L); 3) extract short-channel effect parameters such as Dvt0, Dvt1, Dvt2, Nfactor through one set of devices (large and fixed W different L); 4) extract saturation velocity  parameters such as vsat through one set of devices (large and fixed W$\&$ different L); 5) extract bulk effect parameters such as a0, ags, Keta through one set of devices (large and fixed W$\&$ different L)\cite{Y.Cheng1995}. The model parameters are shown in Table \ref{table2}. The root-mean-square (RMS) error is introduced to estimate the deviation between the results from the measurements and simulations. The RMS error is given by equation (\ref{eq1}) where N is the number of data,  I $_{mi}$ is measured data and I$_{si}$ is simulated data.
\begin{equation}
\mathrm { RMS\;ERROR} = \sqrt { \frac { 1 } { N } \sum _ { i = 1 } ^ { n } \left( \frac { I _ { m i } - I _ { s i } } { I _ { t h } } \right) ^ { 2 } } \times 100
\label{eq1}
\end{equation}
The value of the threshold current I$_{th}$ can be set appropriately to obtain meaningful results. In this case, I$_{th}$ has been set to the maximum measured value according to BSIMProPlus. After the parameters have been changed appropriately, the RMS error between the simulation results and the test data improves to around 6\%(Fig. \ref{fig5}(e-h),Table \ref{table3}).

However, deviations caused by the kink effect still exist in some devices (Fig. \ref{fig6}(a,b,e,f,i,j)) at LHT. We connected a resistor in series with the substrate to improve the fitting precision as shown in Fig. \ref{fig6}(p), the MOSFET represents the BSIM model with cryogenic parameters and the resistor represents the freeze-out effect in the LDD region and substrate. The resistor value was extracted via Matlab using a polunomial fitting method. The characteristic of the non-linear resistor is shown in Fig.\ref{fig6}(m,n,o). Fig. \ref{fig6}(c,d,g,h,k,l) show the corrected  sub-curcuit simulation results. Good agreement with the DC measurements was achieved for devices over the entire voltage range at 4.2K as the RMS errors are summarized in Table \ref{table3}.

\section{Conclusion}

A cryogenic study of SMIC 0.18$\mu$m 1.8V/5V CMOS technology down to 4.2K has been presented. We performed a relatively simple apparatus for executing the cryogenic measurements. A compact model based on BSIM3v3 has been proposed to optimize the deviation between measurement results and simulations using default parameters. V$_{th}$, Ion to Ioff ratio, G$_{m}$max and resistors on chip were tested down to 4.2K. Using the resulting database and SPICE model, we can design and simulate integrated circuits for cryogenic applications including quantum computer readout and control systems.


\section{Acknowledgment}

This work was supported in part by the National Key Research and Development Program of China (Grant No. 2016YFA0301700), In part by the National Natural Science Foundation of China (Grants No.11625419), in part by the Anhui initiative in Quantum information Technologies (Grants No.AHY080000) and this work was partially carried out at the USTC Center for Micro and Nanoscale Research and Fabrication. The authors would like to thank SMIC for devices fabrication and software support.

\section*{References}
\bibliography{mybibfile}

\providecommand{\newblock}{}
\begin{thebibliography}{10}
\expandafter\ifx\csname url\endcsname\relax
  \def\url#1{{\tt #1}}\fi
\expandafter\ifx\csname urlprefix\endcsname\relax\def\urlprefix{URL }\fi
\providecommand{\eprint}[2][]{\url{#2}}

\bibitem{7999541}
Varizat L, Sou G, Mansour M, Alison D and Rhouni A 2017 A low temperature
  0.35$\mu$m cmos technology bsim3.3 model for space instrumentation:
  Application to a voltage reference design {\em 2017 IEEE International
  Workshop on Metrology for AeroSpace (MetroAeroSpace)\/} pp 74--78

\bibitem{604075}
Hsieh C~C, Wu C~Y and Sun T~P 1997 {\em IEEE Journal of Solid-State Circuits\/}
  {\bf 32} 1192--1199 ISSN 0018-9200

\bibitem{4089134}
Yuce M~R, Liu W, Damiano J, Bharath B, Franzon P~D and Dogan N~S 2007 {\em IEEE
  Transactions on Circuits and Systems I: Regular Papers\/} {\bf 54} 420--431
  ISSN 1549-8328

\bibitem{5373940}
Ekanayake S~R, Lehmann T, Dzurak A~S, Clark R~G and Brawley A 2010 {\em IEEE
  Transactions on Electron Devices\/} {\bf 57} 539--547 ISSN 0018-9383

\bibitem{PhysRevApplied.3.024010}
Hornibrook J~M, Colless J~I, Conway~Lamb I~D, Pauka S~J, Lu H, Gossard A~C,
  Watson J~D, Gardner G~C, Fallahi S, Manfra M~J and Reilly D~J 2015 {\em Phys.
  Rev. Applied\/} {\bf 3}(2) 024010
  \urlprefix\url{https://link.aps.org/doi/10.1103/PhysRevApplied.3.024010}

\bibitem{mine:isscc_2017_cryoCMOS}
Charbon E, Sebastiano F, Babaie M, Vladimirescu A, Shahmohammadi M, Staszewski
  R~B, Homulle H~A~R, Patra B, van Dijk J~P~G, Incandela R~M, Song L and
  Valizadehpasha B 2017 Cryo-cmos circuits and systems for scalable quantum
  computing {\em 2017 IEEE International Solid-State Circuits Conference
  (ISSCC)\/} pp 264--265

\bibitem{doi:10.1063/1.4979611}
Homulle H, Visser S, Patra B, Ferrari G, Prati E, Sebastiano F and Charbon E
  2017 {\em Review of Scientific Instruments\/} {\bf 88} 045103
  (\textit{Preprint} \eprint{https://doi.org/10.1063/1.4979611})
  \urlprefix\url{https://doi.org/10.1063/1.4979611}

\bibitem{8358026}
Fu X, Rol M~A, Bultink C~C, van Someren J, Khammassi N, Ashraf I, Vermeulen
  R~F~L, de~Sterke J~C, Vlothuizen W~J, Schouten R~N, Almudever C~G, DiCarlo L
  and Bertels K 2018 {\em IEEE Micro\/} {\bf 38} 40--47 ISSN 0272-1732

\bibitem{doi:10.1021/acs.nanolett.5b02400}
Deng G~W, Wei D, Li S~X, Johansson J~R, Kong W~C, Li H~O, Cao G, Xiao M, Guo
  G~C, Nori F, Jiang H~W and Guo G~P 2015 {\em Nano Letters\/} {\bf 15}
  6620--6625 pMID: 26327140 (\textit{Preprint}
  \eprint{https://doi.org/10.1021/acs.nanolett.5b02400})
  \urlprefix\url{https://doi.org/10.1021/acs.nanolett.5b02400}

\bibitem{PhysRevApplied.9.034011}
Salath\'e Y, Kurpiers P, Karg T, Lang C, Andersen C~K, Akin A, Krinner S,
  Eichler C and Wallraff A 2018 {\em Phys. Rev. Applied\/} {\bf 9}(3) 034011
  \urlprefix\url{https://link.aps.org/doi/10.1103/PhysRevApplied.9.034011}

\bibitem{1484563}
Schrankler J~W, Huang J~S~T, Lutze R~S~L, Vyas H~P and Kirchner G~D 1984
  Cryogenic behavior of scaled cmos devices {\em 1984 International Electron
  Devices Meeting\/} pp 598--600

\bibitem{30958}
Hairapetian A, Gitlin D and Viswanathan C~R 1989 {\em IEEE Transactions on
  Electron Devices\/} {\bf 36} 1448--1455 ISSN 0018-9383

\bibitem{1580601}
Colinge J~, Quinn A~J, Floyd L, Redmond G, Alderman J~C, Xiong W, Cleavelin
  C~R, Schulz T, Schruefer K, Knoblinger G and Patruno P 2006 {\em IEEE
  Electron Device Letters\/} {\bf 27} 120--122 ISSN 0741-3106

\bibitem{Y.Cheng1995}
Cheng Y, Chan M, Hui K, Jeng M, Liu Z, Huang J, Chen K, Chen J, Tu R, Ko P~K
  and Hu C 1995 Bsim3v3 manual \url{http://bsim.berkeley.edu/models/bsim3/}

\bibitem{47796}
Hafez I~M, Ghibaudo G and Balestra F 1990 {\em IEEE Transactions on Electron
  Devices\/} {\bf 37} 818--821 ISSN 0018-9383

\bibitem{BALESTRA19941967}
Balestra F and Ghibaudo G 1994 {\em Solid-State Electronics\/} {\bf 37} 1967 --
  1975 ISSN 0038-1101
  \urlprefix\url{http://www.sciencedirect.com/science/article/pii/0038110194900647}

\bibitem{24362}
Simoen E, Dierickx B, Warmerdam L, Vermeiren J and Claeys C 1989 {\em IEEE
  Transactions on Electron Devices\/} {\bf 36} 1155--1161 ISSN 0018-9383

\bibitem{BALESTRA1987321}
Balestra F, Audaire L and Lucas C 1987 {\em Solid-State Electronics\/} {\bf 30}
  321 -- 327 ISSN 0038-1101
  \urlprefix\url{http://www.sciencedirect.com/science/article/pii/0038110187901900}

\bibitem{3372}
Dierickx B, Warmerdam L, Simoen E~R, Vermeiren J and Claeys C 1988 {\em IEEE
  Transactions on Electron Devices\/} {\bf 35} 1120--1125 ISSN 0018-9383

\bibitem{1052555}
Hanamura H, Aoki M, Masuhara T, Minato O, Sakai Y and Hayashida T 1986 {\em
  IEEE Journal of Solid-State Circuits\/} {\bf 21} 484--490 ISSN 0018-9200

\bibitem{8269294}
Homulle H, Song L, Charbon E and Sebastiano F 2018 {\em IEEE Journal of the
  Electron Devices Society\/} {\bf 6} 263--270 ISSN 2168-6734

\bibitem{8066592}
Beckers A, Jazaeri F, Ruffino A, Bruschini C, Baschirotto A and Enz C 2017
  Cryogenic characterization of 28 nm bulk cmos technology for quantum
  computing {\em 2017 47th European Solid-State Device Research Conference
  (ESSDERC)\/} pp 62--65 ISSN 2378-6558

\bibitem{d0d52351b8e84fcc8e961be5fa144ba2}
Shin M, Shi M, Mouis M, Cros A, Josse E, Kim G~T and Ghibaudo G 2014 Low
  temperature characterization of 14nm fdsoi cmos devices {\em 2014 11th
  International Workshop on Low Temperature Electronics, WOLTE 2014\/} (IEEE
  Computer Society) pp 29--32 ISBN 9781479948420

\bibitem{ZHAO201449}
Zhao H and Liu X 2014 {\em Cryogenics\/} {\bf 59} 49 -- 59 ISSN 0011-2275
  \urlprefix\url{http://www.sciencedirect.com/science/article/pii/S0011227513000969}

\bibitem{1742-6596-834-1-012002}
Varizat L, Sou G and Mansour M 2017 {\em Journal of Physics: Conference
  Series\/} {\bf 834} 012002
  \urlprefix\url{http://stacks.iop.org/1742-6596/834/i=1/a=012002}

\bibitem{8329135}
Incandela R~M, Song L, Homulle H, Charbon E, Vladimirescu A and Sebastiano F
  2018 {\em IEEE Journal of the Electron Devices Society\/} {\bf 6} 996--1006
  ISSN 2168-6734

\bibitem{8424046}
Beckers A, Jazaeri F and Enz C 2018 {\em IEEE Transactions on Electron
  Devices\/} {\bf 65} 3617--3625 ISSN 0018-9383

\bibitem{doi:10.1080/14786436708227694}
Holwech I and Jeppesen J 1967 {\em The Philosophical Magazine: A Journal of
  Theoretical Experimental and Applied Physics\/} {\bf 15} 217--228
  (\textit{Preprint} \eprint{https://doi.org/10.1080/14786436708227694})
  \urlprefix\url{https://doi.org/10.1080/14786436708227694}

\bibitem{doi:10.1063/1.555614}
Matula R~A 1979 {\em Journal of Physical and Chemical Reference Data\/} {\bf 8}
  1147--1298 (\textit{Preprint} \eprint{https://doi.org/10.1063/1.555614})
  \urlprefix\url{https://doi.org/10.1063/1.555614}

\end{thebibliography}

\end{document}